\documentclass[aps, reprint,floatfix,prl]{revtex4-1}

\usepackage[utf8]{inputenc}
\usepackage{t1enc} 
\usepackage{graphicx}   
\usepackage{amsmath}
\usepackage{multirow}

\begin{document}

\title{Co- and contra-directional vertical coupling between ferromagnetic layers with grating for short-wavelength spin wave generation}
\author{Piotr Graczyk}
\email{graczyk@amu.edu.pl}
\author{Mateusz Zelent}
\author{Maciej Krawczyk}
\email{krawczyk@amu.edu.pl}
\affiliation{Faculty of Physics, Adam Mickiewicz University in Poznan, Umultowska 85, 61-614 Poznan, Poland}

\begin{abstract}

The possibility to generate short spin waves is of great interest in the field of magnonics nowadays. We present an effective and technically affordable way of conversion of long spin waves, which may be generated by conventional microwave antenna, to the short, sub-micrometer waves. It is achieved by grating-assisted resonant dynamic dipolar interaction between two ferromagnetic layers separated by some distance. We analyze criteria for the optimal conversion giving a semi-analytical approach for the coupling coefficient. We show by the numerical calculations the efficient energy transfer between layers which may be either of co-directional or contra-directional type. Such a system may operate either as a short spin wave generator or a frequency filter, moving foreward possible application of magnonics.

\end{abstract}

\maketitle


Smaller is better. This paradigm is a driving force in many modern branches of applied physics and technology. The advances in miniaturization of silicon-based logic devices have followed famous Moore’s law  to the 1990s \cite{StanfordElectronicsLab}. Now, the dimensions and operating voltages in CMOS technology are pushed to the limits. Therefore, new ideas for memory and logic devices are desirable and some have already been emerged \cite{Theis2016}.

Some of the novel concepts come from magnonics, which is  promising for processing information field of modern magnetism \cite{Kruglyak10b,RoadMap2014,Grundler_2016}. Magnonics deals with the spin waves (SWs) as a carrier of information and thus it is expected to combine high frequency and short wavelength together with low power consumption \cite{Khitun2010,Lenk2011,Chumak2015}. The possibility of miniaturization of spin-wave-based logic devices is dependent on the wavelength, thus the attention is focused on searching the ways of generation of short SWs.  Several approaches has been recently reported: shortening the wavelength in a tapered waveguide \cite{Demidov2011}, microwave-to-SW transducers based either on magnetic wires \cite{Au2012}, nanostructure edges \cite{Davies2016}, diffraction gratings \cite{Yu2013,Sklenar2012,Yu2016_a}, coplanar waveguides \cite{Maendl_2017} or anisotropy-modulated multiferroic heterostructures \cite{Hamalainen2017}, transduction of acoustic waves into  short SWs due to magneto-elastic coupling \cite{Graczyk2017,Graczyk2017b}, and emitters based on current-driven oscillating pinned domain walls \cite{VandeWiele2016}.

The SW waveguides, multiplexers \cite{Davies_2015}, coupled waveguides with or without exploitation of  periodic gratings \cite{Sadovnikov_2016,Sadovnikov_2015,Sadovnikov_2017b}  have been investigated as basic units for processing information, some of them  were demonstrated experimentally. Most of these studies deal with the in-plane SW propagation. but control of SW propagation in out-of-plane direction is promising for applications, although an area of 3D magnonics still remain unexplored \cite{Fernadez2017}. Multilayer structures have already been studied for controlling SW spectra in thin films with using periodic gratings \cite{Morozova_16,Morozova_16b, Mruczkiewicz2017}, periodic gratings were also used to excite and detect SWs propagating in homogeneous film beneath it. However, the way of transferring of the SW energy in the vertical direction has not yet been established.

In this paper we investigate the effect of grating-assisted directional coupling between SWs propagating in two ferromagnetic layers.  We formulate phenomenological expression for the coupling strength and discuss a co-directional and contra-directional couplings between the layers, which allow for a transfer energy between them. The conditions for efficient resonant coupling are defined. In numerical simulations we consider permalloy (Py)–-CoFeB bilayer separated by nonmagnetic spacer, with Py layer decorated with a grating. We show, that it is possible to achieve complete SW power exchange between the layers, if the structure is optimized for resonant, phase-matched interaction of a proper magnitude. We propose such a system to be a simple and efficient transducer for generation of short SWs in Py from long SWs in CoFeB, which allows to transfer the SW signals in the vertical direction, and to exploit it as a narrow band filters for future magnonic devices.


The phenomenon of coupling and energy exchange between two identical uniform ferromagnetic films has already been  discussed \cite{Sasaki1979,Koike1980} and it was shown with Brillouin light spectroscopy (BLS) for side coupling between two waveguides two years ago \cite{Sadovnikov2015}. Recently, application of that effect, widely used in photonics from many years \cite{Alferness,Tien77}, has been proposed for signal processing with two coupled magnonic waveguides \cite{Wang2017}. In this case, the resonant interaction between SWs propagating in the identical waveguides  exists in relatively wide range of frequencies if the distance between them is properly tuned. If the layers are different, the SWs are decoupled and each wave becomes confined to a single layer. However, the existence of the other layer still affects the dispersion of the SWs in a range of frequencies. We can couple them again by introducing the periodicity of the structure. Folding of the bands to the Brillouin zone will result in multiple crossings of the bands that originate from different layers. Then, by tuning the distance between layers, to ensure  that crossings lies in the region of strong interaction between layers, we can achieve strong hybridization between the SW bands. To sum up, the criteria of efficient power exchange between different ferromagnetic media (waveguides or layers) are the following: (1) the resonance, i.e. the equality of the frequencies, (2) phase matching---equality of the wavevectors, and (3) strong interaction between layers.

\begin{figure}
\includegraphics*[width=0.5\textwidth]{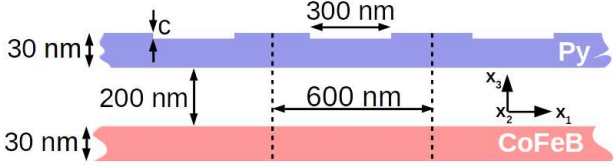}
 \caption{\label{Fig2} The scheme of considered bilayer system composed of CoFeB film and Py film with the periodically corrugated surface. Dashed lines indicate the unit cell.}
\end{figure}

For the efficient energy exchange between SWs in the coupling-in-space mechanism, it is crucial to choose materials with low damping. The best candidates are  yttrium iron garnet (YIG) and CoFeB. However, due to the extremely high magnetic contrast between that materials, it is difficult to match all three conditions: the wavelength of the SW in YIG is much shorter than in CoFeB at the same frequency, and it is the wave mainly driven by the exchange interaction. As a result, the mode in YIG creates very weak stray magnetic field around and the layers must be in an unfeasibly close distance to couple. Due to this, we have chosen Py and CoFeB as materials of significant contrast, but small enough to ensure high stray fields for both SW modes.

Considered bilayer system (Fig.~\ref{Fig2}) consists of CoFeB 30 nm thick homogeneous film and 30 nm Py film. Py is patterned periodically with grooves of 300 nm width and the depth of either 5 nm (Py5) or 15 nm (Py15). The grating has the lattice constant 600 nm. Such dimensions make the structure feasible for fabrication and SWs detection achievable by conventional techniques, such as  BLS or SW spectroscopy. We assume, that the layers are infinite along the $x_2$-coordinate. The separation between layers has been optimized for the sufficient strong interaction at the points of lowest band crossings (discussed latter) and its value is fixed to 200 nm.

To study coupling in the bilayered structure the Landau-Lifshitz-Gilbert equation has been solved either in frequency-domain (for calculation of the dispersion relations) or in the time-dependent computations, both with the use of COMSOL Multiphysics. The details of calculations are described in Ref.  \onlinecite{Mruczkiewicz2013}. We assumed for the CoFeB (Py) layer the values of the exchange constant of 15 pJ/m (13 pJ/m), of the saturation magnetization 1250 kA/m (860 kA/m) and the damping 0.004 (0.01). The results of finite element method have been verified by the micromagnetic simulations performed with MuMax3 \cite{mumax}, which was also used to estimate effectiveness of the SW transfer between the layers and to study filtering properties.

Before presenting the numerical results, we formulate an  approach for calculation the coupling between SWs propagating in two different ferromagnetic layers, a and b, which is based on the coupled mode theory widely used in electromagnetism \cite{Yariv2003, Zhang2008}. A dynamic magnetization $\vec{m}_a$ of the SW mode in the layer a is under influence of the magnetic stray field $\vec{h}_b$ generated by the SW $\vec{m}_b$ in the layer b. The energy of this interaction is:
\begin{equation}
E_{ba}=\frac{1}{S} \int{\vec{m}_a \cdot \vec{h}_b dS},
\label{eq1}
\end{equation}
where integration is conducted through the area $S$ of the unit cell of the structure (see, Fig.~\ref{Fig2}).
The fields are  normalized according to $S^{-1} \int{\vec{m}_i \cdot \vec{h}_i dS } =1, \text{  where  } \ i= \text{a, b}.$ The power transferred from the layer $b$ to the layer $a$ is given by:
\begin{equation}
\kappa_{ba}^{(\omega)}=\frac{1}{S} \int{\frac{\partial \vec{m}_a^*}{\partial t} \cdot \vec{h}_b dS }=i\omega \frac{1}{S} \int{\vec{m}_a^* \cdot \vec{h}_b dS }.
\label{eq3}
\end{equation}
The coupling coefficient $\kappa_{ab}^{(\omega)}$ which describes energy flow from medium $a$ to  $b$ is defined similarly. This formula allows to estimate coupling strength between two waveguides knowing the distributions of stray fields $\vec{h}_i$ and related SW amplitudes $\vec{m}_i$ of the separated layers. As $\kappa \equiv |\kappa_{ab}|=|\kappa_{ba}|$ is a periodic function of  the coordinate in grating-assisted coupler \cite{Huang1994}, the formula (\ref{eq3}) describes an average value of the coupling coefficient in the unit cell.
\begin{figure}
\includegraphics*[width=0.45\textwidth]{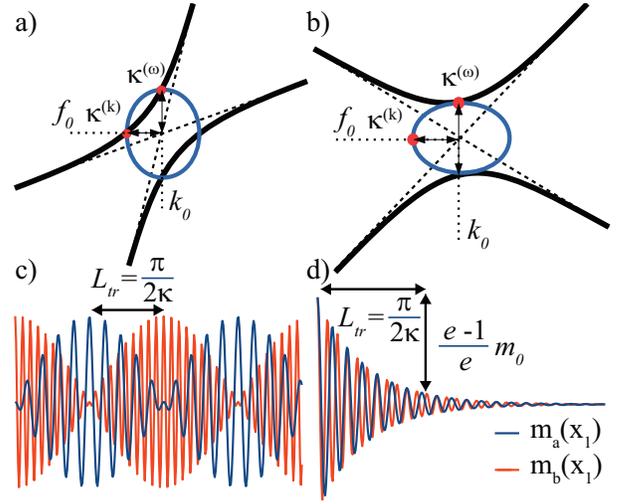}
\caption{\label{Fig1} Possible couplings between SWs in the bilayered structure with periodicity and the definition of characteristic values. (a) Hybridization in co-directional coupling,  (b) hybridization in contra-directional coupling, (c) spatial SW amplitude distribution in the case of co-directional coupling, and (d) spatial amplitude distributions in the case of contra-directional coupling.}
\end{figure}
The value of the coupling coefficient  $\kappa^{(\omega)}$ defines also the width of the band splitting in a dispersion relation in the frequency scale (Fig.~\ref{Fig1}). We can express coupling coefficient also in the wavevector scale by $\kappa^{(k)} = \kappa^{(\omega)}/v$ where $v=(v_a+v_b)/2$, and $v_a$ is the group velocity of SW in the layer $a$ and $v_b$ in the layer $b$. It is assumed that group velocities are constant close to the modes crossing. By this reformulation of the coupling coefficient from the frequency to the wavevector scale, it is possible to describe the coupled modes as a sum of two modes which wavenumbers differ by $\Delta k=2\kappa^{(k)}$ and one of them is symmetric while the other is antisymmetric. 

In the structure with periodicity, the folding back to the first Brillouin zone exists, the two family of bands, these related to SWs in a and b layers, cross each other. The crossing bands can have the same  (Fig.~\ref{Fig1}a) or opposite (Fig.~\ref{Fig1}b) sign of the group velocity. In the first case we call them co-directional and in the second contra-directional couplings. Considering co-directional coupling (Fig.~\ref{Fig1}a), the interference of the two waves which differ by $\Delta k$ and propagate in both layers results in the effect of beating with the spatial frequency defined by $\kappa^{(k)}$ (Fig.~\ref{Fig1}c). In case of contra-directional coupling (Fig.~\ref{Fig1}b), the value of $\kappa^{(k)}$ becomes imaginary at the vicinity of crossing. The wave is a sum of two modes $k_a=k_0+i\kappa$ and $k_b=-k_0+i\kappa$ which propagate in opposite directions. If we suppose that initially the input wave exists only at the layer $a$, the resulting effect is a Bragg reflection: the amplitude of the wave $m_a(x_1,t)$ is a decaying function of $x_1$ (Fig.~\ref{Fig1}d) and the outgoing wave $m_b(x_1,t)$ propagates in the opposite direction.

\begin{figure}
\includegraphics*[width=0.45\textwidth]{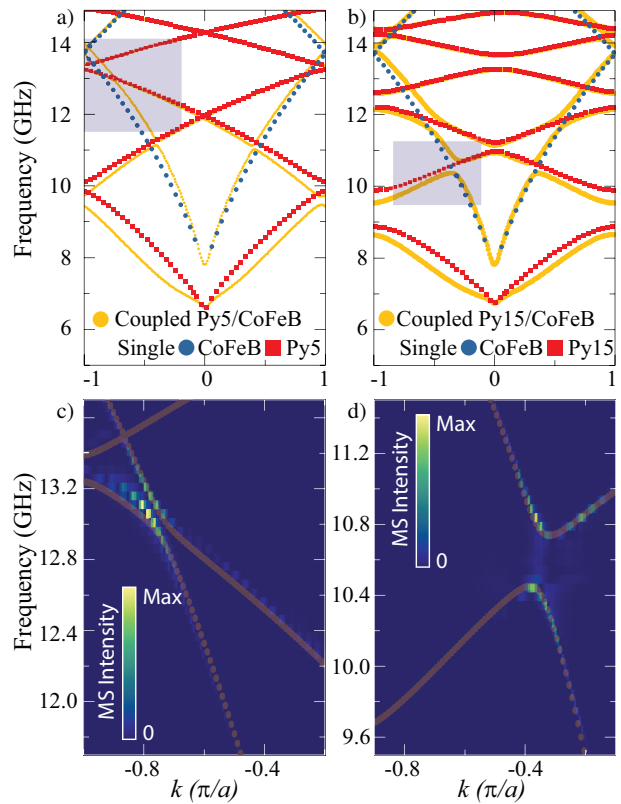}
 \caption{\label{Fig3} Dispersion relations of (a) Py5 (5 nm deep groves)/CoFeB and (b) Py15/CoFeB bilayers (yellow dots) from COMSOL simulations. Dispersions of single Py5, Py15 and CoFeB layers are also indicated. In (c) and (d) the dispersion at the vicinity of the analyzed crossings are shown and superimposed on the results of micormagnetic simulations (color map).}
\end{figure}

Figure~\ref{Fig3}(a) shows the dispersion relation for the CoFeB/Py5 system with (see the yellow dots). Blue and red dots indicate dispersion relations of the isolated Py film with 5 nm deep groves (Py5) and CoFeB films, respectively. It is seen from the plot, that the dispersion of bilayered structure  differ from  the dispersions of isolated films at a range of frequencies.  For $k$ close to zero the frequencies for CoFeB/Py5 are the same as for separate films, because the stray fields are absent for homogeneous excitation and weak in long-wavelength limit. On the other hand, for high values of $k$ dispersion reaches exchange regime and stray fields become also weak. Between that two limits there exists a region of strong interaction.

The effect of nonreciprocity is also apparent from Fig.~\ref{Fig3}(a) for the bilayered structure. It is manifested both, by a difference in the frequency of the waves at +$k$ and -$k$, as well as the differences in the coupling strength on both sides of the Brillouin zone center.  Clearly, hybridization at crossings for the $-k$ are stronger than for the positive values of $k$. These effects are due to nonreciprocity of the Damon-Eshbach mode and the asymmetry of the structure \cite{Mruczkiewicz2017,Henry2017}. 

The branches which originate from the dispersions of two different films anti-crosses at some $k$ points. Thus, choosing appropriate crossing points in the region of strong interaction, we can fulfill the criteria of resonance and phase matching for that waves, required for SW power exchange between the layers. For the time-domain investigation described below we have chosen crossings at 13 GHz (for co-directional coupling) and 10.6 GHz (for contra-directional coupling) presented in Figs. 3(c) and 3(d), respectively.

\begin{figure}
\includegraphics*[width=0.46\textwidth]{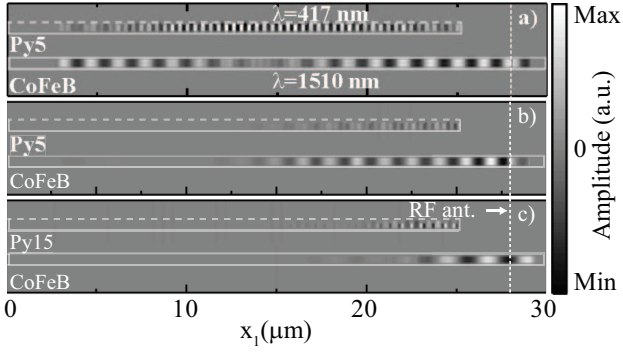}
 \caption{\label{Fig4} Spatial distribution of $m_1$ after 15 ns of sinusoidal excitation by RF antenna in: (a) Py5/CoFeB without damping, (b) Py5/CoFeB with damping, (c) Py15/CoFeB with damping. Lines indicate boundaries of the layers, dashed line indicates periodic surface of Py layer.}
\end{figure}

The line source of the wave in simulations is placed at $x_1 =28$ $\mu$m in the CoFeB film. CoFeB film exists along the whole $x_1$-range, while  Py spreads out from $x_1=0$ to $x_1=25$ $\mu$m. The corrugation of Py film (stripe hole depth) is here $c=5$ nm (Py5). Strong damping close to the both ends of CoFeB and left end of Py is introduced to prevent reflections. Fig.~\ref{Fig4}(a) shows magnetization component $m_1$ after 15 ns of excitation. The amplitude of the SW which goes to the left  is decreasing from the maximum value at  $x_1>25$ $\mu$m to almost zero at $x_1 =14.5$ $\mu$m. Its wavelength is $1.51$ $\mu$m. At the same distance, the SW in Py appears to reach maximum amplitude at $x_1 =14.5$ $\mu$m. Therefore, the energy of the wave in CoFeB has been vertically transformed into the energy of the wave in Py5 completely at a distance of $10.4$ $\mu$m. Wave induced in Py has the wavelength of 417 nm, that is almost 4 times shorter than those which excites it.

The coupling coefficient read out from the dispersion relation [Fig.~\ref{Fig3}(a)] is $\kappa^{(k)}=0.131$ $\mu\text{m}^{-1}$ which gives the value of transfer length $L_{\text{tr}}=\pi/2\kappa^{(k)}=11.9 \ \mu\text{m}$, it is little longer than obtained from time-dependent simulations. The same coupling coefficient calculated from Eq.~(\ref{eq3}) gives the value of 0.145 $\mu\text{m}^{-1}$ and thus $L_{\text{tr}} =10.9$ $\mu$m. 

When the damping is introduced, the effect of energy exchange between layers is strongly suppressed. As it is shown in  Fig.~\ref{Fig4}(b), the short wave in Py5 is still excited by the wave in CoFeB, but it is not accumulated in the layer. The damping length for Py film is of about 2-3 $\mu$m, thus it is much smaller than the length at which transfer from CoFeB occurs. At a given distance, Py film looses more energy than it gets from CoFeB and accumulation cannot take place. 

We can recover accumulation of energy in Py layer with damping by increasing coupling coefficient between layers. For this purpose we increase the corrugation depth of Py to $c=15$ nm (Py15). The dispersion relation for CoFeB/Py15 is shown in Fig. 3(b). It is clearly seen, that the separation between branches increased significantly at the poins of crossings, which indicates increase of coupling strength. The calculated value of $\kappa^{(k)}$ [Eqs.~(\ref{eq1}) and (\ref{eq1})] is 0.43 $\mu\text{m}^{-1}$ which gives $L_{\text{tr}} =3.7$ $\mu$m.

The time-dependent simulations are presented in Fig.~\ref{Fig4}(c). The wave in CoFeB vanish much faster with the distance, because it transfers the energy to Py in shorter time. The excited wave in Py grows and the amplitude reaches maximum value at $x_1 =23$ $\mu$m. Therefore, the energy is clearly accumulated in Py despite damping. At further distance it is seen that small amount of energy is transferred back to CoFeB and again back to Py15.


The crossing shown in Fig.~\ref{Fig1}(b) from the dispersion relation in Fig.~\ref{Fig3}(b) at 10.6 GHz represents contra-directional coupling between SWs in CoFeB and Py of 15 nm deep grooves. The coupling strength estimated from dispersion relation is 0.476 $\mu\text{m}^{-1}$, while the calculated from the Eq.~(\ref{eq3}) based on the stray field amplitude and the magnetization amplitude is 0.435 $\mu\text{m}^{-1}$, that gives the transfer lengths of 3.34 $\mu$m and 3.6 $\mu$m, respectively. 

We performed time domain simulations with the 10.6 GHz excitation at the source line in the system similar to the discussed above for co-directional coupling. The difference is that we introduced the fragment of the homogeneous Py layer at the $x_1$ between 20 $\mu$m and 25 $\mu$m with the strong damping close to the right edge. This allows us to observe the wave going out of the periodic structure toward the right direction.

\begin{figure}
\includegraphics*[width=0.46\textwidth]{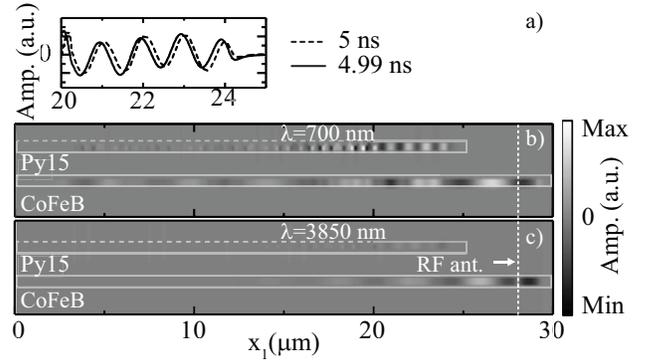}
 \caption{\label{Fig5} Spatial distribution of $m_1$ after 5 ns of sinusoidal excitation of 10.6 GHz by antenna (marked with the yellow line) in: (a) Py15/CoFeB without damping, (b) Py15/CoFeB with damping. Lines indicate boundaries of the layers, dashed line indicate periodic surface of Py layer. (c) $m_1$ in the two consecutive time intervals in the region of uniform Py layer.}
\end{figure}

Figure~\ref{Fig5}(b) shows the results of simulations after 5 ns of excitation. Clearly, the wave of the wavelength of about 700 nm is excited in Py15 by the wave of the wavelength 3.85 $\mu$m in CoFeB. However, the amplitudes of both waves decay with the distance. The distance at which amplitude decreases by the factor 1/e is about 4 $\mu$m. At the output (homogeneous part of Py) we observe the wave which propagates to the right, as shown in Fig.~\ref{Fig5}a. Therefore, the energy of the wave in CoFeB which propagates to the left has been vertically transferred to the energy of the wave in Py, which propagates to the right.

With the damping introduced (Fig.~\ref{Fig5}b), the amplitudes of both waves in Py15 and CoFeB decrease significantly. The amplitude of the outgoing wave in Py is also severely suppressed. However, it is still visible.

Finally, we analyzed with MuMax3 the transmission of the broadband signal (wave packet) excited in CoFeB to the Py layer. The results are shown in Figs. 3(c) and (d) by the color map. As indicated by intensity map, the signal transmits between the layers only at the vicinity of the crossings. That suggests, a system may operate not only as a short-wavelength SW generator, but also as a frequency filter.

In conclusion, we have shown complete resonant energy exchange between the two spin Damon-Eshbach modes propagating in different ferromagnetic layers. It is possible to utilize this effect to excite short SWs from the long SWs or to filter particular resonant frequency from the wave packet. The other interesting option would be to transfer the whole wave packet from the one medium to another. This could be achieved in finite-size waveguides by optimizing their dispersion relation to obtain broadband resonance.

\begin{acknowledgments}
The study has received financial support from the National Science Centre of Poland under grants UMO-2012/07/E/ST3/00538.
\end{acknowledgments}

\bibliography{bibliography}

\begin{thebibliography}{39}%
\makeatletter
\providecommand \@ifxundefined [1]{%
 \@ifx{#1\undefined}
}%
\providecommand \@ifnum [1]{%
 \ifnum #1\expandafter \@firstoftwo
 \else \expandafter \@secondoftwo
 \fi
}%
\providecommand \@ifx [1]{%
 \ifx #1\expandafter \@firstoftwo
 \else \expandafter \@secondoftwo
 \fi
}%
\providecommand \natexlab [1]{#1}%
\providecommand \enquote  [1]{``#1''}%
\providecommand \bibnamefont  [1]{#1}%
\providecommand \bibfnamefont [1]{#1}%
\providecommand \citenamefont [1]{#1}%
\providecommand \href@noop [0]{\@secondoftwo}%
\providecommand \href [0]{\begingroup \@sanitize@url \@href}%
\providecommand \@href[1]{\@@startlink{#1}\@@href}%
\providecommand \@@href[1]{\endgroup#1\@@endlink}%
\providecommand \@sanitize@url [0]{\catcode `\\12\catcode `\$12\catcode
  `\&12\catcode `\#12\catcode `\^12\catcode `\_12\catcode `\%12\relax}%
\providecommand \@@startlink[1]{}%
\providecommand \@@endlink[0]{}%
\providecommand \url  [0]{\begingroup\@sanitize@url \@url }%
\providecommand \@url [1]{\endgroup\@href {#1}{\urlprefix }}%
\providecommand \urlprefix  [0]{URL }%
\providecommand \Eprint [0]{\href }%
\providecommand \doibase [0]{http://dx.doi.org/}%
\providecommand \selectlanguage [0]{\@gobble}%
\providecommand \bibinfo  [0]{\@secondoftwo}%
\providecommand \bibfield  [0]{\@secondoftwo}%
\providecommand \translation [1]{[#1]}%
\providecommand \BibitemOpen [0]{}%
\providecommand \bibitemStop [0]{}%
\providecommand \bibitemNoStop [0]{.\EOS\space}%
\providecommand \EOS [0]{\spacefactor3000\relax}%
\providecommand \BibitemShut  [1]{\csname bibitem#1\endcsname}%
\let\auto@bib@innerbib\@empty
\bibitem [{\citenamefont {{Stanford Electronics
  Lab}}()}]{StanfordElectronicsLab}%
  \BibitemOpen
  \bibfield  {author} {\bibinfo {author} {\bibnamefont {{Stanford Electronics
  Lab}}},\ }\href {https://nano.stanford.edu/cmos-technology-scaling-trend}
  {\enquote {\bibinfo {title} {{CMOS Technology Scaling Trend}},}\
  }\BibitemShut {NoStop}%
\bibitem [{\citenamefont {Theis}\ and\ \citenamefont {Wong}(2016)}]{Theis2016}%
  \BibitemOpen
  \bibfield  {author} {\bibinfo {author} {\bibfnamefont {T.~N.}\ \bibnamefont
  {Theis}}\ and\ \bibinfo {author} {\bibfnamefont {H.-S.~P.}\ \bibnamefont
  {Wong}},\ }\href {\doibase 10.1109/MCSE.2017.29} {\bibfield  {journal}
  {\bibinfo  {journal} {Computing in Science {\&} Engineering}\ }\textbf
  {\bibinfo {volume} {19}},\ \bibinfo {pages} {41} (\bibinfo {year}
  {2016})}\BibitemShut {NoStop}%
\bibitem [{\citenamefont {Kruglyak}\ \emph {et~al.}(2010)\citenamefont
  {Kruglyak}, \citenamefont {Demokritov},\ and\ \citenamefont
  {Grundler}}]{Kruglyak10b}%
  \BibitemOpen
  \bibfield  {author} {\bibinfo {author} {\bibfnamefont {V.~V.}\ \bibnamefont
  {Kruglyak}}, \bibinfo {author} {\bibfnamefont {S.~O.}\ \bibnamefont
  {Demokritov}}, \ and\ \bibinfo {author} {\bibfnamefont {D.}~\bibnamefont
  {Grundler}},\ }\href@noop {} {\bibfield  {journal} {\bibinfo  {journal}
  {Journal of Physics D: Applied Physics}\ }\textbf {\bibinfo {volume} {43}},\
  \bibinfo {pages} {264001} (\bibinfo {year} {2010})}\BibitemShut {NoStop}%
\bibitem [{\citenamefont {Stamps}\ \emph {et~al.}(2014)\citenamefont {Stamps},
  \citenamefont {Breitkreutz}, \citenamefont {{\AA}kerman}, \citenamefont
  {Chumak}, \citenamefont {Otani}, \citenamefont {Bauer}, \citenamefont
  {Thiele}, \citenamefont {Bowen}, \citenamefont {Majetich}, \citenamefont
  {Kl{\"{a}}ui}, \citenamefont {Prejbeanu}, \citenamefont {Dieny},
  \citenamefont {Dempsey},\ and\ \citenamefont {Hillebrands}}]{RoadMap2014}%
  \BibitemOpen
  \bibfield  {author} {\bibinfo {author} {\bibfnamefont {R.~L.}\ \bibnamefont
  {Stamps}}, \bibinfo {author} {\bibfnamefont {S.}~\bibnamefont {Breitkreutz}},
  \bibinfo {author} {\bibfnamefont {J.}~\bibnamefont {{\AA}kerman}}, \bibinfo
  {author} {\bibfnamefont {A.~V.}\ \bibnamefont {Chumak}}, \bibinfo {author}
  {\bibfnamefont {Y.}~\bibnamefont {Otani}}, \bibinfo {author} {\bibfnamefont
  {G.~E.~W.}\ \bibnamefont {Bauer}}, \bibinfo {author} {\bibfnamefont {J.-U.}\
  \bibnamefont {Thiele}}, \bibinfo {author} {\bibfnamefont {M.}~\bibnamefont
  {Bowen}}, \bibinfo {author} {\bibfnamefont {S.~A.}\ \bibnamefont {Majetich}},
  \bibinfo {author} {\bibfnamefont {M.}~\bibnamefont {Kl{\"{a}}ui}}, \bibinfo
  {author} {\bibfnamefont {I.~L.}\ \bibnamefont {Prejbeanu}}, \bibinfo {author}
  {\bibfnamefont {B.}~\bibnamefont {Dieny}}, \bibinfo {author} {\bibfnamefont
  {N.~M.}\ \bibnamefont {Dempsey}}, \ and\ \bibinfo {author} {\bibfnamefont
  {B.}~\bibnamefont {Hillebrands}},\ }\href
  {http://stacks.iop.org/0022-3727/47/i=33/a=333001} {\bibfield  {journal}
  {\bibinfo  {journal} {Journal of Physics D: Applied Physics}\ }\textbf
  {\bibinfo {volume} {47}},\ \bibinfo {pages} {333001} (\bibinfo {year}
  {2014})}\BibitemShut {NoStop}%
\bibitem [{\citenamefont {Grundler}(2016)}]{Grundler_2016}%
  \BibitemOpen
  \bibfield  {author} {\bibinfo {author} {\bibfnamefont {D.}~\bibnamefont
  {Grundler}},\ }\href {\doibase 10.1038/nnano.2016.16} {\bibfield  {journal}
  {\bibinfo  {journal} {Nature Nanotechnology}\ }\textbf {\bibinfo {volume}
  {11}},\ \bibinfo {pages} {407} (\bibinfo {year} {2016})}\BibitemShut
  {NoStop}%
\bibitem [{\citenamefont {Khitun}\ \emph {et~al.}(2010)\citenamefont {Khitun},
  \citenamefont {Bao},\ and\ \citenamefont {Wang}}]{Khitun2010}%
  \BibitemOpen
  \bibfield  {author} {\bibinfo {author} {\bibfnamefont {A.}~\bibnamefont
  {Khitun}}, \bibinfo {author} {\bibfnamefont {M.}~\bibnamefont {Bao}}, \ and\
  \bibinfo {author} {\bibfnamefont {K.~L.}\ \bibnamefont {Wang}},\ }\href
  {\doibase 10.1088/0022-3727/43/26/264005} {\bibfield  {journal} {\bibinfo
  {journal} {Journal of Physics D: Applied Physics}\ }\textbf {\bibinfo
  {volume} {43}},\ \bibinfo {pages} {264005} (\bibinfo {year}
  {2010})}\BibitemShut {NoStop}%
\bibitem [{\citenamefont {Lenk}\ \emph {et~al.}(2011)\citenamefont {Lenk},
  \citenamefont {Ulrichs}, \citenamefont {Garbs},\ and\ \citenamefont
  {Munzenberg}}]{Lenk2011}%
  \BibitemOpen
  \bibfield  {author} {\bibinfo {author} {\bibfnamefont {B.}~\bibnamefont
  {Lenk}}, \bibinfo {author} {\bibfnamefont {H.}~\bibnamefont {Ulrichs}},
  \bibinfo {author} {\bibfnamefont {F.}~\bibnamefont {Garbs}}, \ and\ \bibinfo
  {author} {\bibfnamefont {M.}~\bibnamefont {Munzenberg}},\ }\href {\doibase
  10.1016/j.physrep.2011.06.003} {\bibfield  {journal} {\bibinfo  {journal}
  {Physics Reports}\ }\textbf {\bibinfo {volume} {507}},\ \bibinfo {pages}
  {107} (\bibinfo {year} {2011})},\ \Eprint {http://arxiv.org/abs/1101.0479}
  {arXiv:1101.0479} \BibitemShut {NoStop}%
\bibitem [{\citenamefont {Chumak}\ \emph {et~al.}(2015)\citenamefont {Chumak},
  \citenamefont {Vasyuchka}, \citenamefont {Serga},\ and\ \citenamefont
  {Hillebrands}}]{Chumak2015}%
  \BibitemOpen
  \bibfield  {author} {\bibinfo {author} {\bibfnamefont {A.~V.}\ \bibnamefont
  {Chumak}}, \bibinfo {author} {\bibfnamefont {V.~I.}\ \bibnamefont
  {Vasyuchka}}, \bibinfo {author} {\bibfnamefont {A.~A.}\ \bibnamefont
  {Serga}}, \ and\ \bibinfo {author} {\bibfnamefont {B.}~\bibnamefont
  {Hillebrands}},\ }\href@noop {} {\bibfield  {journal} {\bibinfo  {journal}
  {Nature Physics}\ }\textbf {\bibinfo {volume} {11}},\ \bibinfo {pages} {453}
  (\bibinfo {year} {2015})}\BibitemShut {NoStop}%
\bibitem [{\citenamefont {Demidov}\ \emph {et~al.}(2011)\citenamefont
  {Demidov}, \citenamefont {Kostylev}, \citenamefont {Rott}, \citenamefont
  {Mnchenberger}, \citenamefont {Reiss},\ and\ \citenamefont
  {Demokritov}}]{Demidov2011}%
  \BibitemOpen
  \bibfield  {author} {\bibinfo {author} {\bibfnamefont {V.~E.}\ \bibnamefont
  {Demidov}}, \bibinfo {author} {\bibfnamefont {M.~P.}\ \bibnamefont
  {Kostylev}}, \bibinfo {author} {\bibfnamefont {K.}~\bibnamefont {Rott}},
  \bibinfo {author} {\bibfnamefont {J.}~\bibnamefont {Mnchenberger}}, \bibinfo
  {author} {\bibfnamefont {G.}~\bibnamefont {Reiss}}, \ and\ \bibinfo {author}
  {\bibfnamefont {S.~O.}\ \bibnamefont {Demokritov}},\ }\href {\doibase
  10.1063/1.3631756} {\bibfield  {journal} {\bibinfo  {journal} {Applied
  Physics Letters}\ }\textbf {\bibinfo {volume} {99}},\ \bibinfo {pages}
  {082507} (\bibinfo {year} {2011})}\BibitemShut {NoStop}%
\bibitem [{\citenamefont {Au}\ \emph {et~al.}(2012)\citenamefont {Au},
  \citenamefont {Ahmad}, \citenamefont {Dmytriiev}, \citenamefont {Dvornik},
  \citenamefont {Davison},\ and\ \citenamefont {Kruglyak}}]{Au2012}%
  \BibitemOpen
  \bibfield  {author} {\bibinfo {author} {\bibfnamefont {Y.}~\bibnamefont
  {Au}}, \bibinfo {author} {\bibfnamefont {E.}~\bibnamefont {Ahmad}}, \bibinfo
  {author} {\bibfnamefont {O.}~\bibnamefont {Dmytriiev}}, \bibinfo {author}
  {\bibfnamefont {M.}~\bibnamefont {Dvornik}}, \bibinfo {author} {\bibfnamefont
  {T.}~\bibnamefont {Davison}}, \ and\ \bibinfo {author} {\bibfnamefont
  {V.~V.}\ \bibnamefont {Kruglyak}},\ }\href {\doibase 10.1063/1.4711039}
  {\bibfield  {journal} {\bibinfo  {journal} {Applied Physics Letters}\
  }\textbf {\bibinfo {volume} {100}},\ \bibinfo {pages} {182404} (\bibinfo
  {year} {2012})}\BibitemShut {NoStop}%
\bibitem [{\citenamefont {Davies}\ and\ \citenamefont
  {Kruglyak}(2016)}]{Davies2016}%
  \BibitemOpen
  \bibfield  {author} {\bibinfo {author} {\bibfnamefont {C.~S.}\ \bibnamefont
  {Davies}}\ and\ \bibinfo {author} {\bibfnamefont {V.~V.}\ \bibnamefont
  {Kruglyak}},\ }\href {\doibase 10.1109/TMAG.2016.2517000} {\bibfield
  {journal} {\bibinfo  {journal} {IEEE Transactions on Magnetics}\ }\textbf
  {\bibinfo {volume} {52}},\ \bibinfo {pages} {7378955} (\bibinfo {year}
  {2016})}\BibitemShut {NoStop}%
\bibitem [{\citenamefont {Yu}\ \emph {et~al.}(2013)\citenamefont {Yu},
  \citenamefont {Duerr}, \citenamefont {Huber}, \citenamefont {Bahr},
  \citenamefont {Schwarze}, \citenamefont {Brandl},\ and\ \citenamefont
  {Grundler}}]{Yu2013}%
  \BibitemOpen
  \bibfield  {author} {\bibinfo {author} {\bibfnamefont {H.}~\bibnamefont
  {Yu}}, \bibinfo {author} {\bibfnamefont {G.}~\bibnamefont {Duerr}}, \bibinfo
  {author} {\bibfnamefont {R.}~\bibnamefont {Huber}}, \bibinfo {author}
  {\bibfnamefont {M.}~\bibnamefont {Bahr}}, \bibinfo {author} {\bibfnamefont
  {T.}~\bibnamefont {Schwarze}}, \bibinfo {author} {\bibfnamefont
  {F.}~\bibnamefont {Brandl}}, \ and\ \bibinfo {author} {\bibfnamefont
  {D.}~\bibnamefont {Grundler}},\ }\href {\doibase 10.1038/ncomms3702}
  {\bibfield  {journal} {\bibinfo  {journal} {Nature Communications}\ }\textbf
  {\bibinfo {volume} {4}},\ \bibinfo {pages} {2702} (\bibinfo {year}
  {2013})}\BibitemShut {NoStop}%
\bibitem [{\citenamefont {Sklenar}\ \emph {et~al.}(2012)\citenamefont
  {Sklenar}, \citenamefont {Bhat}, \citenamefont {Tsai}, \citenamefont
  {Delong},\ and\ \citenamefont {Ketterson}}]{Sklenar2012}%
  \BibitemOpen
  \bibfield  {author} {\bibinfo {author} {\bibfnamefont {J.}~\bibnamefont
  {Sklenar}}, \bibinfo {author} {\bibfnamefont {V.~S.}\ \bibnamefont {Bhat}},
  \bibinfo {author} {\bibfnamefont {C.~C.}\ \bibnamefont {Tsai}}, \bibinfo
  {author} {\bibfnamefont {L.~E.}\ \bibnamefont {Delong}}, \ and\ \bibinfo
  {author} {\bibfnamefont {J.~B.}\ \bibnamefont {Ketterson}},\ }\href {\doibase
  10.1063/1.4737438} {\bibfield  {journal} {\bibinfo  {journal} {Applied
  Physics Letters}\ }\textbf {\bibinfo {volume} {101}},\ \bibinfo {pages}
  {052404} (\bibinfo {year} {2012})}\BibitemShut {NoStop}%
\bibitem [{\citenamefont {Yu}\ \emph {et~al.}(2016)\citenamefont {Yu},
  \citenamefont {{d' Allivy Kelly}}, \citenamefont {Cros}, \citenamefont
  {Bernard}, \citenamefont {Bortolotti}, \citenamefont {Anane}, \citenamefont
  {Brandl}, \citenamefont {Heimbach},\ and\ \citenamefont
  {Grundler}}]{Yu2016_a}%
  \BibitemOpen
  \bibfield  {author} {\bibinfo {author} {\bibfnamefont {H.}~\bibnamefont
  {Yu}}, \bibinfo {author} {\bibfnamefont {O.}~\bibnamefont {{d' Allivy
  Kelly}}}, \bibinfo {author} {\bibfnamefont {V.}~\bibnamefont {Cros}},
  \bibinfo {author} {\bibfnamefont {R.}~\bibnamefont {Bernard}}, \bibinfo
  {author} {\bibfnamefont {P.}~\bibnamefont {Bortolotti}}, \bibinfo {author}
  {\bibfnamefont {A.}~\bibnamefont {Anane}}, \bibinfo {author} {\bibfnamefont
  {F.}~\bibnamefont {Brandl}}, \bibinfo {author} {\bibfnamefont
  {F.}~\bibnamefont {Heimbach}}, \ and\ \bibinfo {author} {\bibfnamefont
  {D.}~\bibnamefont {Grundler}},\ }\href {\doibase 10.1038/ncomms11255}
  {\bibfield  {journal} {\bibinfo  {journal} {Nature Communications}\ }\textbf
  {\bibinfo {volume} {7}},\ \bibinfo {pages} {11255} (\bibinfo {year}
  {2016})}\BibitemShut {NoStop}%
\bibitem [{\citenamefont {Maendl}\ \emph {et~al.}(2017)\citenamefont {Maendl},
  \citenamefont {Stasinopoulos},\ and\ \citenamefont {Grundler}}]{Maendl_2017}%
  \BibitemOpen
  \bibfield  {author} {\bibinfo {author} {\bibfnamefont {S.}~\bibnamefont
  {Maendl}}, \bibinfo {author} {\bibfnamefont {I.}~\bibnamefont
  {Stasinopoulos}}, \ and\ \bibinfo {author} {\bibfnamefont {D.}~\bibnamefont
  {Grundler}},\ }\href {\doibase 10.1063/1.4991520} {\bibfield  {journal}
  {\bibinfo  {journal} {Applied Physics Letters}\ }\textbf {\bibinfo {volume}
  {111}},\ \bibinfo {pages} {12403} (\bibinfo {year} {2017})}\BibitemShut
  {NoStop}%
\bibitem [{\citenamefont {H{\"{a}}m{\"{a}}l{\"{a}}inen}\ \emph
  {et~al.}(2017)\citenamefont {H{\"{a}}m{\"{a}}l{\"{a}}inen}, \citenamefont
  {Brandl}, \citenamefont {Franke}, \citenamefont {Grundler},\ and\
  \citenamefont {{Van Dijken}}}]{Hamalainen2017}%
  \BibitemOpen
  \bibfield  {author} {\bibinfo {author} {\bibfnamefont {S.~J.}\ \bibnamefont
  {H{\"{a}}m{\"{a}}l{\"{a}}inen}}, \bibinfo {author} {\bibfnamefont
  {F.}~\bibnamefont {Brandl}}, \bibinfo {author} {\bibfnamefont {K.~J.}\
  \bibnamefont {Franke}}, \bibinfo {author} {\bibfnamefont {D.}~\bibnamefont
  {Grundler}}, \ and\ \bibinfo {author} {\bibfnamefont {S.}~\bibnamefont {{Van
  Dijken}}},\ }\href {\doibase 10.1103/PhysRevApplied.8.014020} {\bibfield
  {journal} {\bibinfo  {journal} {Physical Review Applied}\ }\textbf {\bibinfo
  {volume} {8}},\ \bibinfo {pages} {014020} (\bibinfo {year}
  {2017})}\BibitemShut {NoStop}%
\bibitem [{\citenamefont {Graczyk}\ \emph {et~al.}(2017)\citenamefont
  {Graczyk}, \citenamefont {K{\l}os},\ and\ \citenamefont
  {Krawczyk}}]{Graczyk2017}%
  \BibitemOpen
  \bibfield  {author} {\bibinfo {author} {\bibfnamefont {P.}~\bibnamefont
  {Graczyk}}, \bibinfo {author} {\bibfnamefont {J.}~\bibnamefont {K{\l}os}}, \
  and\ \bibinfo {author} {\bibfnamefont {M.}~\bibnamefont {Krawczyk}},\ }\href
  {\doibase 10.1103/PhysRevB.95.104425} {\bibfield  {journal} {\bibinfo
  {journal} {Physical Review B}\ }\textbf {\bibinfo {volume} {95}},\ \bibinfo
  {pages} {104425} (\bibinfo {year} {2017})}\BibitemShut {NoStop}%
\bibitem [{\citenamefont {Graczyk}\ and\ \citenamefont
  {Krawczyk}(2017)}]{Graczyk2017b}%
  \BibitemOpen
  \bibfield  {author} {\bibinfo {author} {\bibfnamefont {P.}~\bibnamefont
  {Graczyk}}\ and\ \bibinfo {author} {\bibfnamefont {M.}~\bibnamefont
  {Krawczyk}},\ }\href {\doibase 10.1103/PhysRevB.96.024407} {\bibfield
  {journal} {\bibinfo  {journal} {Physical Review B}\ }\textbf {\bibinfo
  {volume} {96}},\ \bibinfo {pages} {024407} (\bibinfo {year} {2017})},\
  \Eprint {http://arxiv.org/abs/1704.06118} {arXiv:1704.06118} \BibitemShut
  {NoStop}%
\bibitem [{\citenamefont {{Van de Wiele}}\ \emph {et~al.}(2016)\citenamefont
  {{Van de Wiele}}, \citenamefont {H{\"{a}}m{\"{a}}l{\"{a}}inen}, \citenamefont
  {Bal{\'{a}}{\v{z}}}, \citenamefont {Montoncello},\ and\ \citenamefont {van
  Dijken}}]{VandeWiele2016}%
  \BibitemOpen
  \bibfield  {author} {\bibinfo {author} {\bibfnamefont {B.}~\bibnamefont {{Van
  de Wiele}}}, \bibinfo {author} {\bibfnamefont {S.~J.}\ \bibnamefont
  {H{\"{a}}m{\"{a}}l{\"{a}}inen}}, \bibinfo {author} {\bibfnamefont
  {P.}~\bibnamefont {Bal{\'{a}}{\v{z}}}}, \bibinfo {author} {\bibfnamefont
  {F.}~\bibnamefont {Montoncello}}, \ and\ \bibinfo {author} {\bibfnamefont
  {S.}~\bibnamefont {van Dijken}},\ }\href {\doibase 10.1038/srep21330}
  {\bibfield  {journal} {\bibinfo  {journal} {Scientific Reports}\ }\textbf
  {\bibinfo {volume} {6}},\ \bibinfo {pages} {21330} (\bibinfo {year}
  {2016})}\BibitemShut {NoStop}%
\bibitem [{\citenamefont {Davies}\ \emph {et~al.}(2015)\citenamefont {Davies},
  \citenamefont {Sadovnikov}, \citenamefont {Grishin}, \citenamefont
  {Sharaevsky}, \citenamefont {Nikitov},\ and\ \citenamefont
  {Kruglyak}}]{Davies_2015}%
  \BibitemOpen
  \bibfield  {author} {\bibinfo {author} {\bibfnamefont {C.~S.}\ \bibnamefont
  {Davies}}, \bibinfo {author} {\bibfnamefont {A.~V.}\ \bibnamefont
  {Sadovnikov}}, \bibinfo {author} {\bibfnamefont {S.~V.}\ \bibnamefont
  {Grishin}}, \bibinfo {author} {\bibfnamefont {Y.~P.}\ \bibnamefont
  {Sharaevsky}}, \bibinfo {author} {\bibfnamefont {S.~A.}\ \bibnamefont
  {Nikitov}}, \ and\ \bibinfo {author} {\bibfnamefont {V.~V.}\ \bibnamefont
  {Kruglyak}},\ }\href {\doibase 10.1109/TMAG.2015.2447010} {\bibfield
  {journal} {\bibinfo  {journal} {IEEE Transactions on Magnetics}\ }\textbf
  {\bibinfo {volume} {51}},\ \bibinfo {pages} {1} (\bibinfo {year}
  {2015})}\BibitemShut {NoStop}%
\bibitem [{\citenamefont {Sadovnikov}\ \emph {et~al.}(2016)\citenamefont
  {Sadovnikov}, \citenamefont {Beginin}, \citenamefont {Morozova},
  \citenamefont {Sharaevskii}, \citenamefont {Grishin}, \citenamefont
  {Sheshukova},\ and\ \citenamefont {Nikitov}}]{Sadovnikov_2016}%
  \BibitemOpen
  \bibfield  {author} {\bibinfo {author} {\bibfnamefont {A.~V.}\ \bibnamefont
  {Sadovnikov}}, \bibinfo {author} {\bibfnamefont {E.~N.}\ \bibnamefont
  {Beginin}}, \bibinfo {author} {\bibfnamefont {M.~A.}\ \bibnamefont
  {Morozova}}, \bibinfo {author} {\bibfnamefont {Y.~P.}\ \bibnamefont
  {Sharaevskii}}, \bibinfo {author} {\bibfnamefont {S.~V.}\ \bibnamefont
  {Grishin}}, \bibinfo {author} {\bibfnamefont {S.~E.}\ \bibnamefont
  {Sheshukova}}, \ and\ \bibinfo {author} {\bibfnamefont {S.~A.}\ \bibnamefont
  {Nikitov}},\ }\href {\doibase 10.1063/1.4960195} {\bibfield  {journal}
  {\bibinfo  {journal} {Applied Physics Letters}\ }\textbf {\bibinfo {volume}
  {109}},\ \bibinfo {pages} {42407} (\bibinfo {year} {2016})}\BibitemShut
  {NoStop}%
\bibitem [{\citenamefont {Sadovnikov}\ \emph
  {et~al.}(2015{\natexlab{a}})\citenamefont {Sadovnikov}, \citenamefont
  {Beginin}, \citenamefont {Sheshukova}, \citenamefont {Romanenko},
  \citenamefont {Sharaevskii},\ and\ \citenamefont
  {Nikitov}}]{Sadovnikov_2015}%
  \BibitemOpen
  \bibfield  {author} {\bibinfo {author} {\bibfnamefont {A.~V.}\ \bibnamefont
  {Sadovnikov}}, \bibinfo {author} {\bibfnamefont {E.~N.}\ \bibnamefont
  {Beginin}}, \bibinfo {author} {\bibfnamefont {S.~E.}\ \bibnamefont
  {Sheshukova}}, \bibinfo {author} {\bibfnamefont {D.~V.}\ \bibnamefont
  {Romanenko}}, \bibinfo {author} {\bibfnamefont {Y.~P.}\ \bibnamefont
  {Sharaevskii}}, \ and\ \bibinfo {author} {\bibfnamefont {S.~A.}\ \bibnamefont
  {Nikitov}},\ }\href {\doibase 10.1063/1.4936207} {\bibfield  {journal}
  {\bibinfo  {journal} {Applied Physics Letters}\ }\textbf {\bibinfo {volume}
  {107}},\ \bibinfo {pages} {202405} (\bibinfo {year}
  {2015}{\natexlab{a}})}\BibitemShut {NoStop}%
\bibitem [{\citenamefont {Sadovnikov}\ \emph {et~al.}(2017)\citenamefont
  {Sadovnikov}, \citenamefont {Odintsov}, \citenamefont {Beginin},
  \citenamefont {Sheshukova}, \citenamefont {Sharaevskii},\ and\ \citenamefont
  {Nikitov}}]{Sadovnikov_2017b}%
  \BibitemOpen
  \bibfield  {author} {\bibinfo {author} {\bibfnamefont {A.~V.}\ \bibnamefont
  {Sadovnikov}}, \bibinfo {author} {\bibfnamefont {S.~A.}\ \bibnamefont
  {Odintsov}}, \bibinfo {author} {\bibfnamefont {E.~N.}\ \bibnamefont
  {Beginin}}, \bibinfo {author} {\bibfnamefont {S.~E.}\ \bibnamefont
  {Sheshukova}}, \bibinfo {author} {\bibfnamefont {Y.~P.}\ \bibnamefont
  {Sharaevskii}}, \ and\ \bibinfo {author} {\bibfnamefont {S.~A.}\ \bibnamefont
  {Nikitov}},\ }\href {\doibase 10.1109/TMAG.2017.2709540} {\bibfield
  {journal} {\bibinfo  {journal} {IEEE Transactions on Magnetics}\ }\textbf
  {\bibinfo {volume} {PP}},\ \bibinfo {pages} {1} (\bibinfo {year}
  {2017})}\BibitemShut {NoStop}%
\bibitem [{\citenamefont {Fern{\'{a}}ndez-Pacheco}\ \emph
  {et~al.}(2017)\citenamefont {Fern{\'{a}}ndez-Pacheco}, \citenamefont
  {Streubel}, \citenamefont {Fruchart}, \citenamefont {Hertel}, \citenamefont
  {Fischer},\ and\ \citenamefont {Cowburn}}]{Fernadez2017}%
  \BibitemOpen
  \bibfield  {author} {\bibinfo {author} {\bibfnamefont {A.}~\bibnamefont
  {Fern{\'{a}}ndez-Pacheco}}, \bibinfo {author} {\bibfnamefont
  {R.}~\bibnamefont {Streubel}}, \bibinfo {author} {\bibfnamefont
  {O.}~\bibnamefont {Fruchart}}, \bibinfo {author} {\bibfnamefont
  {R.}~\bibnamefont {Hertel}}, \bibinfo {author} {\bibfnamefont
  {P.}~\bibnamefont {Fischer}}, \ and\ \bibinfo {author} {\bibfnamefont
  {R.~P.}\ \bibnamefont {Cowburn}},\ }\href {\doibase 10.1038/ncomms15756}
  {\bibfield  {journal} {\bibinfo  {journal} {Nature Communications}\ }\textbf
  {\bibinfo {volume} {8}},\ \bibinfo {pages} {15756} (\bibinfo {year}
  {2017})}\BibitemShut {NoStop}%
\bibitem [{\citenamefont {Morozova}\ \emph
  {et~al.}(2016{\natexlab{a}})\citenamefont {Morozova}, \citenamefont
  {Sharaevskaya}, \citenamefont {Sadovnikov}, \citenamefont {Grishin},
  \citenamefont {Romanenko}, \citenamefont {Beginin}, \citenamefont
  {Sharaevskii},\ and\ \citenamefont {Nikitov}}]{Morozova_16}%
  \BibitemOpen
  \bibfield  {author} {\bibinfo {author} {\bibfnamefont {M.~A.}\ \bibnamefont
  {Morozova}}, \bibinfo {author} {\bibfnamefont {A.~Y.}\ \bibnamefont
  {Sharaevskaya}}, \bibinfo {author} {\bibfnamefont {A.~V.}\ \bibnamefont
  {Sadovnikov}}, \bibinfo {author} {\bibfnamefont {S.~V.}\ \bibnamefont
  {Grishin}}, \bibinfo {author} {\bibfnamefont {D.~V.}\ \bibnamefont
  {Romanenko}}, \bibinfo {author} {\bibfnamefont {E.~N.}\ \bibnamefont
  {Beginin}}, \bibinfo {author} {\bibfnamefont {Y.~P.}\ \bibnamefont
  {Sharaevskii}}, \ and\ \bibinfo {author} {\bibfnamefont {S.~A.}\ \bibnamefont
  {Nikitov}},\ }\href {\doibase 10.1063/1.4971410} {\bibfield  {journal}
  {\bibinfo  {journal} {Journal of Applied Physics}\ }\textbf {\bibinfo
  {volume} {120}},\ \bibinfo {pages} {223901} (\bibinfo {year}
  {2016}{\natexlab{a}})}\BibitemShut {NoStop}%
\bibitem [{\citenamefont {Morozova}\ \emph
  {et~al.}(2016{\natexlab{b}})\citenamefont {Morozova}, \citenamefont
  {Sharaevskaya}, \citenamefont {Matveev}, \citenamefont {Beginin},\ and\
  \citenamefont {Sharaevskii}}]{Morozova_16b}%
  \BibitemOpen
  \bibfield  {author} {\bibinfo {author} {\bibfnamefont {M.~A.}\ \bibnamefont
  {Morozova}}, \bibinfo {author} {\bibfnamefont {A.~Y.}\ \bibnamefont
  {Sharaevskaya}}, \bibinfo {author} {\bibfnamefont {O.~V.}\ \bibnamefont
  {Matveev}}, \bibinfo {author} {\bibfnamefont {E.~N.}\ \bibnamefont
  {Beginin}}, \ and\ \bibinfo {author} {\bibfnamefont {Y.~P.}\ \bibnamefont
  {Sharaevskii}},\ }\href {\doibase 10.3103/S1541308X16010015} {\bibfield
  {journal} {\bibinfo  {journal} {Physics of Wave Phenomena}\ }\textbf
  {\bibinfo {volume} {24}},\ \bibinfo {pages} {1} (\bibinfo {year}
  {2016}{\natexlab{b}})}\BibitemShut {NoStop}%
\bibitem [{\citenamefont {Mruczkiewicz}\ \emph {et~al.}(2017)\citenamefont
  {Mruczkiewicz}, \citenamefont {Graczyk}, \citenamefont {Lupo}, \citenamefont
  {Adeyeye}, \citenamefont {Gubbiotti},\ and\ \citenamefont
  {Krawczyk}}]{Mruczkiewicz2017}%
  \BibitemOpen
  \bibfield  {author} {\bibinfo {author} {\bibfnamefont {M.}~\bibnamefont
  {Mruczkiewicz}}, \bibinfo {author} {\bibfnamefont {P.}~\bibnamefont
  {Graczyk}}, \bibinfo {author} {\bibfnamefont {P.}~\bibnamefont {Lupo}},
  \bibinfo {author} {\bibfnamefont {A.}~\bibnamefont {Adeyeye}}, \bibinfo
  {author} {\bibfnamefont {G.}~\bibnamefont {Gubbiotti}}, \ and\ \bibinfo
  {author} {\bibfnamefont {M.}~\bibnamefont {Krawczyk}},\ }\href {\doibase
  10.1103/PhysRevB.96.104411} {\bibfield  {journal} {\bibinfo  {journal} {Phys.
  Rev. B}\ }\textbf {\bibinfo {volume} {96}},\ \bibinfo {pages} {104411}
  (\bibinfo {year} {2017})}\BibitemShut {NoStop}%
\bibitem [{\citenamefont {Sasaki}\ and\ \citenamefont
  {Mikoshiba}(1979)}]{Sasaki1979}%
  \BibitemOpen
  \bibfield  {author} {\bibinfo {author} {\bibfnamefont {H.}~\bibnamefont
  {Sasaki}}\ and\ \bibinfo {author} {\bibfnamefont {N.}~\bibnamefont
  {Mikoshiba}},\ }\href {\doibase 10.1063/1.329134} {\bibfield  {journal}
  {\bibinfo  {journal} {Electronics Letters}\ }\textbf {\bibinfo {volume}
  {15}},\ \bibinfo {pages} {172} (\bibinfo {year} {1979})}\BibitemShut
  {NoStop}%
\bibitem [{\citenamefont {Koike}(1980)}]{Koike1980}%
  \BibitemOpen
  \bibfield  {author} {\bibinfo {author} {\bibfnamefont {T.}~\bibnamefont
  {Koike}},\ }in\ \href@noop {} {\emph {\bibinfo {booktitle} {Ultrasonics
  Symposium}}}\ (\bibinfo  {publisher} {IEEE},\ \bibinfo {year} {1980})\ pp.\
  \bibinfo {pages} {552--556}\BibitemShut {NoStop}%
\bibitem [{\citenamefont {Sadovnikov}\ \emph
  {et~al.}(2015{\natexlab{b}})\citenamefont {Sadovnikov}, \citenamefont
  {Beginin}, \citenamefont {Sheshukova}, \citenamefont {Romanenko},
  \citenamefont {Sharaevskii},\ and\ \citenamefont {Nikitov}}]{Sadovnikov2015}%
  \BibitemOpen
  \bibfield  {author} {\bibinfo {author} {\bibfnamefont {A.~V.}\ \bibnamefont
  {Sadovnikov}}, \bibinfo {author} {\bibfnamefont {E.~N.}\ \bibnamefont
  {Beginin}}, \bibinfo {author} {\bibfnamefont {S.~E.}\ \bibnamefont
  {Sheshukova}}, \bibinfo {author} {\bibfnamefont {D.~V.}\ \bibnamefont
  {Romanenko}}, \bibinfo {author} {\bibfnamefont {Y.~P.}\ \bibnamefont
  {Sharaevskii}}, \ and\ \bibinfo {author} {\bibfnamefont {S.~A.}\ \bibnamefont
  {Nikitov}},\ }\href {\doibase 10.1063/1.4936207} {\bibfield  {journal}
  {\bibinfo  {journal} {Applied Physics Letters}\ }\textbf {\bibinfo {volume}
  {107}},\ \bibinfo {pages} {202405} (\bibinfo {year}
  {2015}{\natexlab{b}})}\BibitemShut {NoStop}%
\bibitem [{\citenamefont {Alferness}(1981)}]{Alferness}%
  \BibitemOpen
  \bibfield  {author} {\bibinfo {author} {\bibfnamefont {R.}~\bibnamefont
  {Alferness}},\ }\href {\doibase 10.1109/JQE.1981.1071209} {\bibfield
  {journal} {\bibinfo  {journal} {IEEE Journal of Quantum Electronics}\
  }\textbf {\bibinfo {volume} {17}},\ \bibinfo {pages} {946} (\bibinfo {year}
  {1981})}\BibitemShut {NoStop}%
\bibitem [{\citenamefont {Tien}(1977)}]{Tien77}%
  \BibitemOpen
  \bibfield  {author} {\bibinfo {author} {\bibfnamefont {P.~K.}\ \bibnamefont
  {Tien}},\ }\href {\doibase 10.1103/RevModPhys.49.361} {\bibfield  {journal}
  {\bibinfo  {journal} {Rev. Mod. Phys.}\ }\textbf {\bibinfo {volume} {49}},\
  \bibinfo {pages} {361} (\bibinfo {year} {1977})}\BibitemShut {NoStop}%
\bibitem [{\citenamefont {Wang}\ \emph {et~al.}(2017)\citenamefont {Wang},
  \citenamefont {Pirro}, \citenamefont {Verba}, \citenamefont {Slavin},
  \citenamefont {Hillebrands},\ and\ \citenamefont {Chumak}}]{Wang2017}%
  \BibitemOpen
  \bibfield  {author} {\bibinfo {author} {\bibfnamefont {Q.}~\bibnamefont
  {Wang}}, \bibinfo {author} {\bibfnamefont {P.}~\bibnamefont {Pirro}},
  \bibinfo {author} {\bibfnamefont {R.}~\bibnamefont {Verba}}, \bibinfo
  {author} {\bibfnamefont {A.}~\bibnamefont {Slavin}}, \bibinfo {author}
  {\bibfnamefont {B.}~\bibnamefont {Hillebrands}}, \ and\ \bibinfo {author}
  {\bibfnamefont {A.~V.}\ \bibnamefont {Chumak}},\ }\href
  {http://arxiv.org/abs/1704.02255} {\enquote {\bibinfo {title}
  {{Reconfigurable nano-scale spin-wave directional coupler}},}\ } (\bibinfo
  {year} {2017}),\ \Eprint {http://arxiv.org/abs/1704.02255} {arXiv:1704.02255}
  \BibitemShut {NoStop}%
\bibitem [{\citenamefont {Mruczkiewicz}\ \emph {et~al.}(2013)\citenamefont
  {Mruczkiewicz}, \citenamefont {Krawczyk}, \citenamefont {Sakharov},
  \citenamefont {Khivintsev}, \citenamefont {Filimonov},\ and\ \citenamefont
  {Nikitov}}]{Mruczkiewicz2013}%
  \BibitemOpen
  \bibfield  {author} {\bibinfo {author} {\bibfnamefont {M.}~\bibnamefont
  {Mruczkiewicz}}, \bibinfo {author} {\bibfnamefont {M.}~\bibnamefont
  {Krawczyk}}, \bibinfo {author} {\bibfnamefont {V.~K.}\ \bibnamefont
  {Sakharov}}, \bibinfo {author} {\bibfnamefont {Y.~V.}\ \bibnamefont
  {Khivintsev}}, \bibinfo {author} {\bibfnamefont {Y.~A.}\ \bibnamefont
  {Filimonov}}, \ and\ \bibinfo {author} {\bibfnamefont {S.~A.}\ \bibnamefont
  {Nikitov}},\ }\href {\doibase 10.1063/1.4793085} {\bibfield  {journal}
  {\bibinfo  {journal} {Journal of Applied Physics}\ }\textbf {\bibinfo
  {volume} {113}},\ \bibinfo {pages} {093908} (\bibinfo {year}
  {2013})}\BibitemShut {NoStop}%
\bibitem [{\citenamefont {Vansteenkiste}\ \emph {et~al.}(2014)\citenamefont
  {Vansteenkiste}, \citenamefont {Leliaert}, \citenamefont {Dvornik},
  \citenamefont {Helsen}, \citenamefont {Garcia-Sanchez},\ and\ \citenamefont
  {Waeyenberge}}]{mumax}%
  \BibitemOpen
  \bibfield  {author} {\bibinfo {author} {\bibfnamefont {A.}~\bibnamefont
  {Vansteenkiste}}, \bibinfo {author} {\bibfnamefont {J.}~\bibnamefont
  {Leliaert}}, \bibinfo {author} {\bibfnamefont {M.}~\bibnamefont {Dvornik}},
  \bibinfo {author} {\bibfnamefont {M.}~\bibnamefont {Helsen}}, \bibinfo
  {author} {\bibfnamefont {F.}~\bibnamefont {Garcia-Sanchez}}, \ and\ \bibinfo
  {author} {\bibfnamefont {B.~V.}\ \bibnamefont {Waeyenberge}},\ }\href
  {\doibase 10.1063/1.4899186} {\bibfield  {journal} {\bibinfo  {journal} {AIP
  Advances}\ }\textbf {\bibinfo {volume} {4}},\ \bibinfo {pages} {107133}
  (\bibinfo {year} {2014})},\ \Eprint
  {http://arxiv.org/abs/http://dx.doi.org/10.1063/1.4899186}
  {http://dx.doi.org/10.1063/1.4899186} \BibitemShut {NoStop}%
\bibitem [{\citenamefont {Yariv}\ and\ \citenamefont {Yeh}(2003)}]{Yariv2003}%
  \BibitemOpen
  \bibfield  {author} {\bibinfo {author} {\bibfnamefont {A.}~\bibnamefont
  {Yariv}}\ and\ \bibinfo {author} {\bibfnamefont {P.}~\bibnamefont {Yeh}},\
  }\href@noop {} {\emph {\bibinfo {title} {{Optical Waves in Crystals}}}}\
  (\bibinfo  {publisher} {John Wiley {\&} Sons},\ \bibinfo {year} {2003})\ pp.\
  \bibinfo {pages} {177--201}\BibitemShut {NoStop}%
\bibitem [{\citenamefont {Zhang}\ and\ \citenamefont {Li}(2008)}]{Zhang2008}%
  \BibitemOpen
  \bibfield  {author} {\bibinfo {author} {\bibfnamefont {K.}~\bibnamefont
  {Zhang}}\ and\ \bibinfo {author} {\bibfnamefont {D.}~\bibnamefont {Li}},\
  }\href@noop {} {\emph {\bibinfo {title} {{Electromagnetic Theory for
  Microwaves and Optoelectronics}}}}\ (\bibinfo  {publisher}
  {Springer-Verlag},\ \bibinfo {address} {Berlin},\ \bibinfo {year} {2008})\
  pp.\ \bibinfo {pages} {450--461}\BibitemShut {NoStop}%
\bibitem [{\citenamefont {Huang}(1994)}]{Huang1994}%
  \BibitemOpen
  \bibfield  {author} {\bibinfo {author} {\bibfnamefont {W.-P.}\ \bibnamefont
  {Huang}},\ }\href
  {http://www.osapublishing.org/josaa/fulltext.cfm?uri=josaa-11-3-963{\&}id=676}
  {\bibfield  {journal} {\bibinfo  {journal} {Journal of the Optical Society of
  America}\ }\textbf {\bibinfo {volume} {11}},\ \bibinfo {pages} {963}
  (\bibinfo {year} {1994})}\BibitemShut {NoStop}%
\bibitem [{\citenamefont {Henry}\ \emph {et~al.}(2016)\citenamefont {Henry},
  \citenamefont {Gladii},\ and\ \citenamefont {Bailleul}}]{Henry2017}%
  \BibitemOpen
  \bibfield  {author} {\bibinfo {author} {\bibfnamefont {Y.}~\bibnamefont
  {Henry}}, \bibinfo {author} {\bibfnamefont {O.}~\bibnamefont {Gladii}}, \
  and\ \bibinfo {author} {\bibfnamefont {M.}~\bibnamefont {Bailleul}},\
  }\href@noop {} {\bibfield  {journal} {\bibinfo  {journal} {ArXiv e-prints}\ }
  (\bibinfo {year} {2016})},\ \Eprint {http://arxiv.org/abs/1611.06153}
  {arXiv:1611.06153 [cond-mat.mes-hall]} \BibitemShut {NoStop}%
\end{thebibliography}%

\end{document}